\input harvmac
\input epsf

\Title{\vbox{\rightline{BONN-TH-99-03}
\rightline{hep-th/9902068}}
}
{\vbox{\centerline{Membrany corrections}\centerline{to the string anti-string
potential}\centerline{ 
in M5-brane theory }}}
\medskip

\centerline{\it
Stefan F\"orste} 
\bigskip
\centerline{Institute of Physics, Bonn University}
\centerline{Nu\ss allee 12,
Bonn 53115, Germany}
\smallskip

\vglue .3cm
\bigskip

\noindent
We study the potential between a string and an anti-string source in
M5-theory by using the adS/CFT duality conjecture. We find that the
next to leading order corrections in a saddle point approximation
renormalize the classical result.

\Date{}

\overfullrule=0pt
\parskip=0pt plus 1pt
\sequentialequations
\def\ads5{\hbox{\rm AdS}_5}
\def\s5{S^5}
\def\half{{1\over 2}}
\def\del{\partial}

\def\pl{Phys.\ Lett.}
\def\prl{ Phys.\ Rev.\ Lett.}

\def\ap{Ann.\ Phys.}

%
\def\WITTEN{E.\ Witten, {\it Five brane effective action in M-theory},
J.\ Geom.\ Phys.\ {\bf 22} (1997) 103, {\tt hep-th/9610234}.}
\def\SOROKIN{I.\ Bandos, K.\ Lechner, A.\ Nurmagambetov, P.\ Pasti,
D.\ Sorokin and M.\ Tonin, {\it Covariant action for the
superfive-brane of M theory,} Phys.\ Rev.\ Lett.\ {\bf 78} (1997) 4332,
{\tt hep-th/9701149}.}
\def\SCHWARZ{M.\ Aganagic, J.\ Park, C.\ Popescu and J.\ Schwarz, {\it
World volume action of the M-theory five brane.} Nucl.\ Phys.\ {\bf
B496} (1997) 191, {\tt hep-th/9701166}.}
\def\SEIBERG{N.\ Seiberg, {\it New theories in six dimensions and
matrix description of M-theory on $T^5$ and $T^5/Z_2$}, Phys.\ Lett.\
{\bf B408} (1997) 98, {\tt hep-th/9705221}.}
\def\MALDADS{J.\ Maldacena, {\it The large N limit of superconformal 
field theories and supergravity}, Adv.\ Theor.\ Math.\ Phys.\ {\bf 2}
(1998) 231, {\tt hep-th/ 9711200}.} 
\def\GUKLPO{S.\ Gubser, I.\ Klebanov and A.\ Polyakov, {\it Gauge
theory correlators from non-critical string theory}, \pl {\bf B428} (1998) 
105, {\tt hep-th/9802109}.}
\def\WIHOL{E.\ Witten, {\it Anti de Sitter space and holography},
Adv.\ Theor.\ Math.\ Phys.\ {\bf 2} (1998) 253,
{\tt hep-th/9802150}.}
\def\METTS{R.\ Metsaev and A.\ Tseytlin, {\it Type IIB superstring action in 
$\ads5\times\s5$ background}, Nucl.\ Phys. {\bf B533} (1998) 109, {\tt
hep-th/9805028}.} 
\def\MALLOOP{J.\ Maldacena, {\it Wilson loops in large N field theories}, 
\prl {\bf 80} (1998) 4859, {\tt hep-th/9803002}.}
\def\GROSS{D.\ Gross and H.\ Ooguri, {\it Aspects of large N gauge
theory dynamics as seen by string theory}, Phys.\ Rev.\ {\bf D58}
(1998) 106002, {\tt hep-th/9805129}.}
\def\KALLOSH{R.\ Kallosh and A.\ Rajamaran, {\it Vacua of M-theory and
string theory}, Phys.\ Rev.\ {\bf D58} (1998) 125003, {\tt hep-th/9805041}.}
\def\WIT{B.\ de Wit, K.\ Peeters, J.\ Plefka and A.\ Sevrin, {\it The
M-theory two-brane in $AdS_4\times S^7$ and $AdS_7\times S^4$}, Phys.\
Lett.\ {\bf B443} (1998) 153, {\tt hep-th/9808052}.}
\def\AGFRMU{L.\ Alvarez-Gaum\' e, D.\ Freedman and S.\ Mukhi, {\it The 
background field method and the ultraviolet structure of the supersymmetric
nonlinear sigma models},  \ap {\bf 134} (1981) 85.}
\def\CLAUS{P.\ Claus, {\it Super M-brane actions in $adS_4 \times S^7$
and $adS_7\times S^4$,} {\tt hep-th/9809045}.}
\def\GILKEY{T.\ Branson, P.\ Gilkey and D.\ Vassilevich, {\it Vacuum
expectation value asymptotics for second order differential operators
on manifolds with boundary}, J.\ Math.\ Phys.\ {\bf 39} (1998) 1040,
{\tt hep-th/9702178}.} 
\def\US{S.\ F\"orste, D.\ Ghoshal and S.\ Theisen, work in progress .}
\def\OLESEN{J.\ Greensite and P.\ Olesen, {\it World sheet
fluctuations and the heavy quark potential in the adS/CFT approach},
{\tt hep-th/9901057}.}
{\nopagenumbers

\rightline{}
\ \bigskip\bigskip
%

\ftno=0

}

\newsec{Introduction}
One of the outstanding problems in M-theory is a better understanding
of the world volume theory of the M five-brane. In the recent past
there have been various publications discussing that issue from
different points of view. A small sample of references is given in
\ref\Witten{\WITTEN}, \ref\Sorokin{\SOROKIN}, \ref\Schwarz{\SCHWARZ}\ and
\ref\Seiberg{\SEIBERG}. In the present paper we will use two
descriptions of the M-theory five-brane. What we will call a
``perturbative'' description is the picture that the five-brane is
formed by defects in eleven dimensional Minkowski space which arise
due to open membranes ending on flat $5+1$ dimensional
hypersurfaces. From this perspective the world volume theory has
longitudinal and transversal degrees of freedom. Since we will not
enter a quantitative discussion relying on the perturbative
description the qualitative picture will be sufficient for
us. (Nevertheless it should be interesting to study the presented
configuration from an effective field theoretic approach.)

The dual description which we will call ``non-perturbative'' is based
on Maldacena's conjecture \ref\MaldadS{\MALDADS}\ (further elaborated in
\ref\GuKlPo{\GUKLPO}\ref\WiHol{\WIHOL}), where the five-brane theory
is given by M-theory on $adS_7\times S^4$. In the context of this
paper we will read the M of M-theory as an abbreviation for membrane. 
The precise statement is that M-theory on the space with the metric
\eqn\target{
ds^2 = l_p^2 R^{2}\left[ U^2 dx_{\parallel}^2 + 4 {dU^2 \over U^2}
+d\Omega_4^2\right]}
is equivalent to the world volume theory of $N$ M5-branes sitting on
top of each other. The eleven dimensional Planck length has to be
taken to zero in the end. (Since it drops out of all our final
results we will put it formally to one from now on.) In difference to
\MaldadS\ we have rescaled the world-volume coordinates of the
five-brane $x_\parallel \rightarrow R^{3/2} x_\parallel$. The radius R
(in Planck units) is related to the number of five-branes by the
relation
\eqn\radius{
R = \left( \pi N\right)^{1\over 3} .}
The supergravity solution \target\ is reliable for a large number
of five-branes $N$. 

We want to apply this duality to the computation
of a potential energy density between two straight string sources in
the M5-brane theory. In the next section we will do this by a saddle
point approximation. (This has already been discussed in
\ref\Maloop{\MALLOOP}.) In the following sections we will study
corrections to this result due to membrane fluctuations.   

\newsec{The Background}
In the present paper we take as a ``perturbative'' definition of the
M5-brane theory the picture that the degrees of freedom on the world
volume of the M5-brane are described by membranes ending on
them. The term ``perturbative'' means here that the embedding space is
11 dimensional Minkowski space. (This is in analogy to the
perturbative definition of Yang-Mills theory on D-branes.) Especially
we are interested in a situation where we have one M5 brane separated
by a very large distance from a bunch of $N$ M5 branes. In addition we
span two straight Membranes between the five-branes such that they end
in two parallel strings with distance $L$ on the M5-branes (Fig.\ 1).
\bigskip 

\centerline{\epsfbox{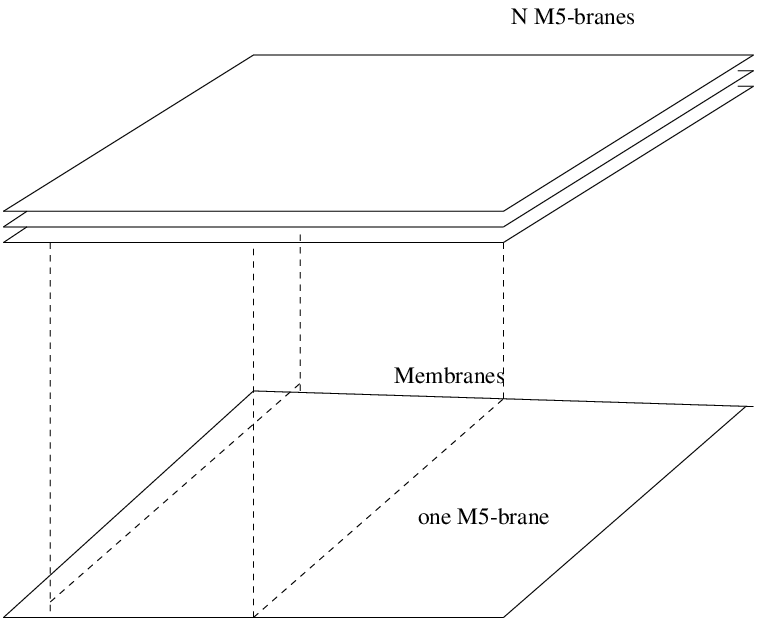}}\nobreak
\centerline{Fig. 1: ``perturbative'' picture:}\centerline{ the
embedding space is 
flat}
\bigskip
 
On the world volume theory of the N M5-branes this corresponds to a
pair of a string anti-string\foot{``Anti'' refers to the opposite
orientation from the five dimensional point of view.}.
The $L$ dependent part of the energy density of the two membranes
corresponds to the potential energy density of the string anti-string
pair in the M5-brane-theory. Since in the perturbative picture the
gravitational interaction in the bulk is neglected the force between
the two membranes is solely carried by M5 world volume fields. 
The
$L$ independent part of the energy density arises due to 
the self energy of the two membranes. It should be proportional to the
separation distance of the single M5-brane from the $N$ M5-branes.
In the M5 field theoretic description longitudinal and transversal
modes couple to the string sources. Exchanges of longitudinal quanta
will lead to the L-dependent potential whereas the transversal quanta
result in an L-independent contribution which diverges when the single
M5 is taken infinitely far away from the $N$ M5-branes. 
Here, we are in the strange situation that we do not know how to
compute this potential energy density in the ``perturbative'' regime
but we do know how to do it non-perturbatively.

The non-perturbative dual of the above configuration is given by a
membrane living in $adS_7\times S^4$ with the boundary condition that
it ends in two parallel strings separated by a distance $L$ at the
boundary of $AdS^7\times S^4$ (Fig.\ 2).
\bigskip

\centerline{\epsfbox{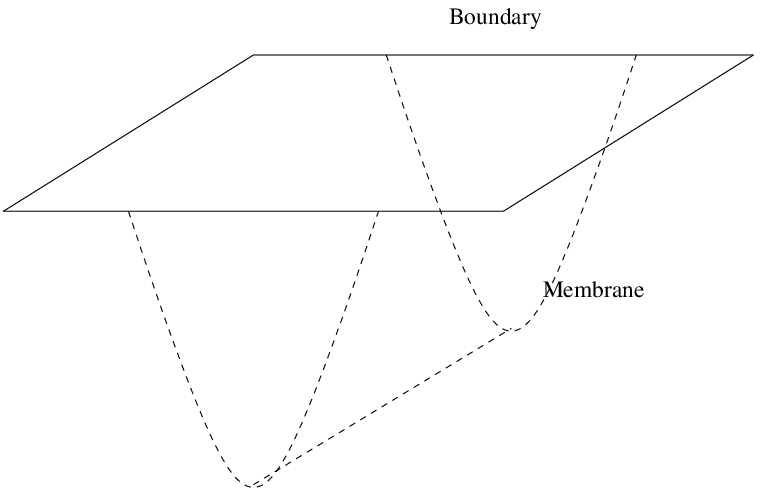}}\nobreak
\centerline{Fig. 2: ``non-perturbative'' picture:}\centerline{ the
embedding space is 
$adS_7\times S^4$}
\bigskip

Since the fermionic zero-modes of the membrane background in
$adS_7\times S^4$ are zero for our problem we can obtain the
configuration of fig.\ 2 by minimizing the world volume of the
membrane which is the Nambu-Goto action
\eqn\Nambu{
S= {1\over 2\pi}
\int d^3\sigma \sqrt{ -\det\left( G_{\mu\nu}\del_a X^\mu \del_b
X^\nu\right)} ,}
where $a=(\tau,\sigma,\phi)$ labels the world volume coordinates of
the membrane 
and $G_{\mu\nu}$ is the metric of the embedding space \target .
We chose a static gauge
\eqn\static{
X^0 = \tau \quad , \quad X^1 = \sigma \quad , \quad X^2 = \phi .}
Further we take the ansatz $U=U\left(\sigma\right)$ and the rest of
the embedding coordinates is constant. Like in the string case one can
employ the explicit $\sigma $ independence of the Lagrangian to reduce
the equations of motion to a first order differential equation
\eqn\first{
\del_\sigma U = \pm {U^2\over 2U_0 ^3}\sqrt{U^6 - U_0^6},}
where $U_0$ is an integration constant which we will relate to the
string anti-string distance $L$.
(In the following we will restrict ourself to positive values of
$\sigma$ and chose the upper sign in \first .) Eq.\ \first\ can be
integrated to give $X^1 = \sigma$ as a function of $U$
\eqn\xone{
X^1 ={2\over U_0}\int\limits_1^{U\over U_0} dy
{1\over y^2\sqrt{y^6 -1}}.}  
The boundary condition $X^1\left(U=\infty\right) = {L\over 2}$ leads
to the relation
\eqn\constant{
U_0 = {2\over 3L}B\left({2\over 3} ,{1 \over 2}\right) ,}
where $B$ denotes Euler's Beta-function.
In order to obtain the energy density we integrate the Lagrange
density in \Nambu\ over $\sigma$ and substitute $\sigma =\sigma\left(U\right)$.
\eqn\energytotal{
\varepsilon = R^3\lim_{U_{max}\to\infty}{2 \over \pi}U_0^2
\int\limits_1^{U_{max}\over U_0} dy {y^4\over \sqrt{y^6-1}},}
where we have introduced an upper cut-off for the $U$ integration.
Now, we split the energy-density into self-energy contribution and a
potential energy density
\eqn\split{
\varepsilon = \varepsilon_{self} +\varepsilon_{pot}.}
with 
\eqn\selfenergy{\eqalign{
\varepsilon_{self} & =R^3 \lim_{U_{max}\to\infty}{1 \over \pi}U_0^2
\int\limits_1^{U_{max}\over U_0} dy {2y^6 + 1 \over y^2
\sqrt{y^6-1}}\cr
& = R^3{U_{max}^2\over \pi}+\ldots ,}}
where the dots stand for terms vanishing in the limit
$U_{max}\to\infty$.
The potential energy density comes out to be
\eqn\potenergy{\eqalign{
\varepsilon_{pot} & = - R^3 {U_0^2\over \pi}\int\limits_1^\infty
dy {1 \over y^2\sqrt{y^6 - 1}}\cr
& = - {2R^3\over 27\pi}B\left({2\over 3} , {1\over 2}\right)^3 {1\over
L^2}.}}
A few remarks are in order. By rescaling $x_\parallel \to
R^{3/2}x_\parallel$ we had changed the world volume of the
M5-brane. To undo this we should divide the energy densities by $R^3$
but at the same time also replace $L\to R^{-3/2}L$\ \foot{These rescalings may
look strange at the first sight but are correct. The energy density is
measured with respect to a volume which is $R^3$ times smaller than the
original one (including the time) and our $L$ has to be expressed in
terms of the original one which is $R^{3/2}$ times bigger.}. This removes the
$R$ dependence of the self energy contribution \selfenergy\ but leaves
the potential energy \potenergy\ unchanged.
Upon compactifying $X^2$ and $\phi$ on a circle (double dimensional
reduction) one obtains a Coulomb
law in a $4+1$ dimensional theory - a result which has been used
already in \ref\Gross{\GROSS}. Even though its derivation is given
in \Maloop\ we decided
to present it in some detail since in the rest of the paper we will
study fluctuations around this background membrane. 

\newsec{Fluctuations}
The result of the previous section is valid for large $R$ where the
supergravity background (geometry) is reliable as well as the saddle
point approximation is good. 
In \ref\Kallosh{\KALLOSH} it was argued that there are no
corrections to the geometry due to finite $N$. Therefore, we will
focus on corrections resulting from fluctuations around the background
membrane. In order to do so we have also to include the fermionic
fluctuations. A $\kappa$-symmetric action for the membrane on
$adS_7\times S^4$ can be found in \ref\Wit{\WIT}. After
rescaling their fermionic coordinates $\theta \to \sqrt{R}\theta $ the
only $R$ dependence of the action appears as an overall factor of
$R^3$. Therefore the loop expansion gives a power series in
$1/R^3$. We will be interested in the next to leading order ($R^0$)
contribution to the potential energy density \potenergy .
To this end, we need to background field expand the membrane action to
second order in fluctuations. Since the background in the $S^4$
direction and in fermionic directions is trivial the bosonic
fluctuations in $adS_7$ direction, in $S^4$ direction and in fermionic
directions decouple and we can discuss their actions separately. In
order to obtain translation invariant functional measures
we use the normal coordinate expansion developed
in \ref\Alv{\AGFRMU}. There fluctuations are parameterized by tangent
vectors $\xi^a$ (with $a$ being a Lorentz index) to geodesics
connecting the background with its fluctuation. It is useful to take
the world volume metric to be the full induced metric because it saves
one from dealing with constraints. With this remarks the calculation
should be straightforward and we will not enter into its details but
just present the results. 

\subsec{Fluctuations in $adS_7$ direction}
The part of the action second order in fluctuations $\xi^a$
($a=0,\ldots ,6$) is 
\eqn\adsfluc{\eqalign{
S_{adS}^{(2)} = & {1\over 4\pi}\int d^3\sigma\sqrt{-h}\left[
h^{ij}\left(\sum\limits_{a=3}^{5} \del_i\xi^a\del_j\xi^a
+\del_i\xi^\perp\del_j\xi^\perp \right)\right.\cr &
\quad \left. +{3\over 4}\sum\limits_{a=3}^5
\left( \xi^a\right)^2 + {3\over 4}\left(1 - {2U_0^6\over
U^6}\right)\left(\xi^\perp\right)^2 \right],}}
where the metric $h_{ij}$ is (up to a factor of $R^2$) the induced
background metric
\eqn\indumet{
ds^2 = - U^2 d\tau^2 + {U^8\over U_0^6}d\sigma^2 + U^2 d\phi^2,}
and we have redefined
\eqn\perpa{\eqalign{
\xi^\parallel &= {U_0^3\over U^3}\xi^1 +{\sqrt{U^6 - U_0^6}\over
U^3}\xi^6 \cr 
\xi^\perp &= -{\sqrt{U^6 - U_0^6}\over U^3}\xi^1 +{U_0^3\over U^3}\xi^6.}}
Note that the Jacobian of this redefinition is one. The new fields
$\xi^\parallel , \xi^\perp $ are fluctuations which lie in the one-six
plane and parameterize fluctuations parallel respectively perpendicular to the
background membrane. 

We observe that \adsfluc\ degenerates since it does not depend on
$\xi^0, \xi^2,$ and $\xi^\parallel$. This originates from the freedom
of performing world volume diffeomorphisms. We remove the degeneracy
by gauge fixing
\eqn\diffix{
\xi^0 =\xi^2 = \xi^\parallel = 0 .}

\subsec{Fluctuations in $S^4$ direction}
Since the background is trivial in the $S^4$ direction we obtain a
very simple action quadratic in $\xi^a$ ($a=7,\ldots ,10$)
\eqn\sfluc{
S_S^{(2)} = {1\over 4\pi}\int d^3\sigma\sqrt{-h}
h^{ij}\sum\limits_{a=7}^{10} \del_i\xi^a\del_j\xi^a .}

\subsec{Fluctuations in fermionic directions}
In order to obtain the part of the action quadratic in fermionic
fluctuations we need to take the part of the membrane action \Wit\ 
bilinear in fermions and plug in there our background for the bosons. 
Then one obtains a result containing only $\Gamma^a$ ($a=0,\ldots,
6$), where $\Gamma^a$ denotes an eleven dimensional Gamma-matrix. 
Now, one can write $\Gamma^a = \gamma^a \otimes \gamma^{5^\prime}$ where
$\gamma^a$ is a gamma-matrix of the seven-dimensional tangent space of
$adS_7$ and $\gamma^{5^\prime}$ belongs to the tangent space of
$S^4$. We split  our 32-component spinor into two sixteen component
spinors $\theta^1,\theta^2$ according to their eigenvalue with
respect to  $\gamma^{5^\prime}$,
\eqn\split{
\gamma^{5^\prime}\theta^1 = \theta^1 \quad , \quad
\gamma^{5^\prime}\theta^2 = - \theta^2 .}
Further we should fix $\kappa$-symmetry. A $\kappa$-fixed action of 
the membrane on $adS_7\times S^4$ is discussed in \ref\Claus{\CLAUS},
for our purpose we find a different gauge fixing convenient however.
First, define (cf \perpa )
\eqn\gammapa{
\gamma^\parallel = {U_0^3\over U^3}\gamma^1 + {\sqrt{U^6 - U_0^6}\over
U^3}\gamma^6.}
With this we choose as a $\kappa$-fixing condition
\eqn\kappafix{\eqalign{
\left( 1 + \gamma^{0\parallel 2}\right) \theta^1 & = 0\cr
\left( 1 - \gamma^{0\parallel 2}\right) \theta^2 & = 0.}}
Further we will need the dreibeine and spin-connections
belonging to \indumet\ (numbers denote Lorentz-indices),
\eqn\dreib{
e^0 _\tau =e^2 _\phi = U \quad ,\quad e^1 _\sigma = {U^4\over
U_0^3}\quad, \quad \omega_\tau ^{01} = \omega_\phi ^{21} = {\sqrt{U^6
- U_0^6} \over 2 U^2},}
and all other components are zero. With some algebra and employing
\kappafix\ one can write the equations of motion for the fermionic
fluctuations as follows
\eqn\eomone{\rho^a e_a ^i\left(\del_i + {1\over 4}\omega_i ^{bc}\rho_b
\rho_c + 
A_i\right)\theta^1 = -{3\over 4}\theta^1 ,}
where we have defined $\rho$-matrices satisfying a $2+1$ dimensional
Clifford algebra 
\eqn\rhomatrix{
\rho^0 = \gamma^0 \quad , \quad \rho^1 = \gamma^{02}\quad , \quad
\rho^2 = \gamma^2 .}
The field
\eqn\gaugefield{
A_\sigma = {3 U\over 4}\gamma^{16} }
appears as a background value of a gauge field belonging to local
rotations in the one-six plane (the $\rho$'s commute with $A$).
For $\theta^2$ we obtain the same equation \eomone\ but with $\rho^1$
replaced by $-\rho^1$.
So, the condition \kappafix\ allows us to write the equations of
motion for the fermionic fluctuations in a covariant three dimensional
form where the target space spinors `metamorphosed' into world-volume
spinors. 

Multiplying the kinetic operator from \eomone\ with its adjoint gives
\eqn\square{
-\Delta - {R^{(3)}\over 4} + {9 \over 16},}
where $R^{(3)}$ is the scalar curvature computed from \indumet\ 
\eqn\curvature{
R^{(3)} = {3 \over 2} {U^6 + U_0^6\over U^6}}
and $\Delta$ is the Laplacian including spin- and gauge connections.
\subsec{Adding up the fluctuations}
From \adsfluc\ \sfluc\ and \square\ we obtain for the one loop
effective action
\eqn\effaction{\eqalign{
S_{eff}^{1-loop} = & \half \log \det{}^3\left( -\Delta_0
+{3\over4}\right) \cr 
 & \quad +\half \log {\det \left( -\Delta_0 +{9\over 4}-R^{(3)} \right)
\det{}^4 \left(-\Delta_0\right) \over \det\left(-\Delta - {1 \over 4}
R^{(3)} + {9\over 16}\right)},}}
where $\Delta_0$ is the Laplacian with respect to \indumet\ acting on scalars.
The power in the fermionic determinant (in the denominator) results
from 32 real fermionic 
components which have been reduced to 16 by $\kappa$-fixing. Since
$\Delta $ is an eight by eight matrix and we have squared the
fermionic operators we arrive at the expression \effaction .
Unfortunately we are not able to evaluate the full expression
\effaction . However, we can extract the uv-divergent contributions.
The formulas we are going to use can be found e.g.\ in
\ref\Gilkey{\GILKEY}. In $2+1$ dimensions there are two potentially
divergent contributions to the effective action. For an operator of
the form $-\Delta + E$ there is a cubic divergence
\eqn\cubic{
a_0 = {1\over \Lambda^3}\left( 4\pi\right)^{-{3\over 2}}\tr {\bf 1}}
where the trace is taken over all fields and includes an integration
with the covariant measure. Since in our case the Laplacian is
dimensionless ($U$ has mass dimension one) the short distance cut-off
$\Lambda$ is dimensionless as well. (When taking the limit of
vanishing Planck length the short distance cut-off in Planck units is
held fixed.) From \effaction\ we see that we do not encounter cubic
divergences. The linear divergence is 
\eqn\linear{
a_2 = {\left(4\pi\right)^{-{3\over 2}}\over
6\Lambda}\tr\left(6E+R^{(3)}\right).}
We have\foot{We take care of the halfs in front of the logarithms in
\effaction\ by
restricting the $\sigma$ integration on the region between zero and $L/2$.}
\eqn\tre{
\tr E = \int d^3 \sigma \sqrt{-h}R^{(3)}}
and hence the divergent contribution to the effective energy density is
\eqn\endiv{
\varepsilon^{div} = {1\over\Lambda}\left(4\pi\right)^{-{3\over
2}}\int d\sigma 
\sqrt{-h} R^{(3)} = \varepsilon^{div}_{self}+\varepsilon^{div}_{pot}.}
For the divergent contribution to the self energy we find
\eqn\selfdiv{
\varepsilon^{div}_{self} = {3 \over 4\Lambda}\left( 4\pi\right)^{-{3\over
2}} \int d\sigma {2U^6 + U_0^6 \over U_0^3} = {3\over
2\Lambda}\left(4\pi\right)^{-{3\over 2}}U_{max}^2 + \ldots ,}
where the dots stand again for terms vanishing when $U_{max}$ is taken
to infinity. The potential energy density receives the following
divergent contribution
\eqn\potdiv{
\varepsilon^{div}_{pot} = {3\over 4\Lambda}\left
( 4\pi\right)^{-{3\over 2}}\int d\sigma U_0^3 = {1\over 9\Lambda}\left
( 4\pi\right)^{- {3\over 2}} B\left({2\over 3},{1\over 2}\right)^3
{1\over L^2}.}
So, both the self energy and the potential energy are renormalized.

So far we did not mention boundary contributions which appear in
theorem 4.1 of \Gilkey . We take Dirichlet boundary conditions (the
configuration on the M5 brane is not allowed to fluctuate). By
fermion-boson matching one realizes that $a_1$ and the boundary
contribution to $a_2$ vanish but $a_3$ does not. This leads to the
following additional part in the energy density
\eqn\beff{
{\log\Lambda\over 16 \pi}\sqrt{-h^{(2)}} R^{(3)}_{| U=U_{max}},}
where $h^{(2)}_{ij}$ is obtained from \indumet\ by fixing
$U=U_{max}$ ($\sigma =constant$). In the limit $U_{max}\to\infty$ one
observes  that \beff\ gives an additional contribution to the self
energy density $\varepsilon^{div}_{self}$. 

To summarize, we have linearly divergent contributions to the self
energy density \selfdiv\ and to the potential energy density \potdiv ,
in addition there is a logarithmically divergence in the self energy
density \beff . Since the self energy density is infinite in our set
up from the beginning their renormalization does not look like a real
problem. (It can be absorbed in a redefinition of the infinite
$U$-integration cut off $U_{max}$.) The part which is difficult to
interpret is the linear divergence \potdiv . It does not introduce an
additional scale since $\Lambda$ is dimensionless. However, our
calculation seems to imply that in the $AdS_7\times S^4$ case the
Maldacena conjecture needs to be supplemented by the information to
what value the UV cut off (in Planck units) has to be taken in the
near horizon limit. We do not know to which data of the M5 brane
theory this information belongs.

\newsec{Conclusions}
Starting from the Maldacena conjecture and assuming that M-theory
is described by membranes we computed the potential energy density
between a string and an anti-string source in the M5-brane theory.
We found that after double dimensional reduction the potential energy
follows a $4+1$ dimensional Coulomb law. Then we discussed corrections
to this result due to membrane fluctuations to the next to leading order.
We proposed a way of world volume diffeomorphism and $\kappa$-symmetry
fixing which seems suitable for the given problem. Finally, we found
that the next to leading order renormalizes the classical result.

Analogous techniques can be applied to the Wilson loop computation
based on the duality between ${\cal N} =4$ supersymmetric Yang-Mills
theory and string theory on $adS_5\times S^5$ \Maloop . A publication dealing
with that case is in preparation \ref\Us{\US}. But even without going
through the details of the calculation one can guess what to
expect. According to \ref\Mets{\METTS}\ divergent contributions to the
partition function of the string on $adS_5\times S^5$ result at most
in a constant contribution to the dilaton beta-function. Hence, a
potential divergence will arise with a factor $\int \sqrt{-h}R^{(2)}$
where $h$ is now the induced metric resulting from the background
discussed in \Maloop. Computing this integral one finds that it only
contains a self energy contribution (in the case of $D3$ branes one
can associate this to a $W$ mass via the Higgs mechanism).

It should be interesting to extend the presented discussion to the
finite temperature case. There it may be possible to deduce in certain
limits more than just the divergent contributions to the partition
function
(cf \ref\Ole{\OLESEN}). 
\bigskip

\centerline{\bf Acknowledgments}
This work is supported in part by the EC programmes
ERB-FMRX-CT-96-0090 and ERB-FMRX-CT-96-0045. 
I would like to thank Debashis Ghoshal and Stefan Theisen for
collaboration on \Us\ where related issues are discussed.
Further I acknowledge discussions with Harald Dorn, Kasper Peeters and
Max Zucker.  
\listrefs \bye